\documentclass[12pt]{article}
\usepackage{graphicx}
\usepackage{enumerate}
\usepackage{amsmath}
\usepackage{amsthm}
\usepackage{amsfonts}
\usepackage{amssymb}
\usepackage{tikz}
\usepackage{sgame}
\usepackage{mathtools}
\usepackage[affil-it]{authblk}
\newcommand{\bt}{\begin{theorem}}

	\newcommand{\ds}{\displaystyle}
	\newcommand{\et}{\end{theorem}}
\newcommand {\be}{\begin{equation}}
\newcommand {\ee}{\end{equation}}
\newcommand{\vsp}{\vskip 1em}

\newcommand{\ol}{\overline}
\newcommand{\noi}{\noindent}

\newcommand{\wh}{\widehat}
\def \qed {\hfill \vrule height6pt width6pt depth0pt}

\title{Partition function form games with probabilistic beliefs}
\begin{document}

	\date{\today}
	
	\author{\it{Paraskevas Lekeas}\footnote{Morton Grove, IL, USA; plekeas@gmail.com}\\ Giorgos Stamatopoulos\footnote{Department of Economics, University of Crete, Rethymno, Crete, Greece; gstamato@uoc.gr}}

	\maketitle
	
\begin{abstract}\noi We revisit games in partition function form, i.e. cooperative games where the payoff of a coalition depends on the partition of the entire set of players. We assume that each coalition computes its worth having probabilistic beliefs over the coalitional behavior of the outsiders, i.e., it assigns various probability distributions over the set of partitions that the outsiders can form.  These beliefs are not necessarily consistent with respect to the actual choices of the outsiders.
We apply this framework to symmetric partition function form games characterized by either positive or negative externalities and we derive conditions on coalitional beliefs that guarantee the non-emptiness of the core of the induced games.
	\end{abstract}
	
\vspace{0.1cm} \noi\hspace{0.83cm} {\small{{\it{Keywords:}} partition function; positive externalities; probabilistic belief; core
			
\vspace{0.15cm} \noi\hspace{0.83cm} \noi \it{JEL Classification:}} C71}
	
	\section{Introduction}
	Cooperative game theory studies situations where coalitions of players act collectively by signing binding agreements. One of the starting points of the theory is to determine the worth a coalition can achieve. In games with orthogonal coalitions, i.e., coalitions that do not affect one another, this task is quite straightforward as it suffices to study the actions of the members of  the coalition at hand only. In this case the worth is captured by the characteristic function, which assigns a single number to each coalition (assuming utility is transferable).
	When orthogonality is absent, or in other words, when there are inter-coalitional externalities, the specification of the worth of a coalition requires the studying of the actions of the outside players, and in particular their coalition structure. This situation is modeled via the partition function (Thrall and Lucas 1963), which assigns a number to a pair consisting of a coalition and the partition in which the coalition is embedded.
	
	The most important solution concept of cooperative games is the core, which lists all the ways to allocate the societal surplus that are immune to coalitional rejections. In characteristic function form games the core is unambiguously defined, due to the lack of externalities. However when dealing with partition function form games, the core is not uniquely defined as it depends on the conjectures a deviant coalition has about the behavior of the non-members. According to \textit{$\alpha$ and $\beta$-conjectures} (Aumann 1959), the members of a coalition compute their worth assuming that the outside players are interested in  minimizing the payoff of the coalition. According to \textit{$\gamma$-conjectures} (Hart $\&$ Kurz 1983; Chander $\&$ Tulkens 1997), it is assumed that the outsiders form only singleton coalitions. On the other extreme, \textit{$\delta$-conjectures} presume that the outsiders form a single coalition (Hart $\&$ Kurz 1983). Another line of research looks again on certain partitions of the outsiders but it posits that the deviant coalition assumes for itself some sort of Stackelberg leadership in the game (Currarini $\&$ Marini 2003, Marini 2013).
	
The \textit{r-approach} (Huang $\&$ Sjostrom 2003; K\'{o}czy 2007) proposes that the members of a coalition compute their worth by looking recursively on the sub-games played among the outsiders. On a more recent approach, it is assumed that the expectations of a coalition regarding the reaction of the outsiders are guided by a set of axioms (Bloch $\&$ Nouweland 2014). Finally, the study of conjectures has been extended to include situations with multiple spheres of externalities  (Nax 2014). Externalities in a certain sphere are borne out of inter-coalitional interactions within that sphere (intra-sphere externalities) or from interactions in other spheres (inter-sphere externalities).\footnote{An extended survey of approaches, results, applications, etc. of partition function form games can be found in K\'{o}czy (2018).}

The current paper  takes a different route. It assumes that when a group of players, $S$, contemplate to break off from a partition (in our case the partition consisting of the grand coalition), they assign various probability distributions on the set of all possible partitions.  These probabilistic beliefs do not necessarily reflect the behavior of the outsiders, i.e., beliefs need not be consistent with actual choices. Given the beliefs, no matter how they form, one can compute the expected worth of $S$ and define the core of the resulting cooperative game.  The task that arises then is to find conditions on the data of the game that guarantee the non-emptiness of the core. 

We apply our framework to symmetric partition function form games. We mainly focus on the case of positive externalities from coalition formation (Hafalir 2007) but we also discuss the case of negative externalities (due to the similarity of the analysis). We show the non-emptiness of the core using an induction argument on the number of players in the game.  To do so, we impose certain restrictions on how the probabilistic beliefs of a coalition evolve as the number of players in the game changes. In a nutshell, as the number of players in the game increases these beliefs give relatively lower probability weight to partitions consisting of few members (under positive externalities); or relatively lower weight to partitions consisting of many members (under negative externalities).

We note that Lekeas $\&$ Stamatopoulos (2014) and Stamatopoulos (2021) analyze two applications of probabilistic coalitional beliefs for the case where externalities are generated within an oligopolistic framework (in the former paper) or an economy with environmental externalities (in the latter paper). In particular, Lekeas $\&$ Stamatopoulos (2014) examines cooperative games generated in Cournot markets where the beliefs of a coalition are represented by a specific probability distribution, the logit distribution. Stamatopoulos (2021) examines an economy where the agents' production of goods generates multilateral environmental externalities. When forming a coalition, its members have probabilistic beliefs over the coalitional behavior of the non-members, which are termed admissible beliefs. Actually the current paper builds on admissible beliefs and extends the analysis of Stamatopoulos (2021) to general (but still symmetric) environments.
	
The motivation of our paper is twofold. First, we are interested in generalizing (some of)  the existing theoretical approaches on the definition of core. For example, the notion of $\gamma$-core (which corresponds to $\gamma$-conjectures) is a special case of our approach that arises when each coalition assigns probability one to the event that the outsiders form only singleton coalitions; likewise, the notion of $\delta$-core (which corresponds to $\delta$-conjectures) arises when each coalition assigns probability one to the event that the outsiders form one coalition. Secondly, our paper could be read as a work on bounded rationality in relation to cooperative games. The assignment of various, ad hoc to some extent, probability distributions on the set of partitions of the outsiders may reflect the cognitive inability of the members of a coalition to accurately deduce the outsiders' equilibrium partition. In this sense, probabilistic beliefs act as a rule of thumb. This approach is particularly relevant for games with a large number of players, where the number of different partitions is very large.

\section{Preliminaries}	
Consider a set of players $N=\{1,2,\ldots,n\}$. A partition of $N$ is a collection of disjoint subsets (coalitions) of $N$ whose union is $N$. Let $\pi$ be a such a partition and let $\Pi$ be the set of all partitions. A  partition function specifies the worth of each coalition given the partition in which it is embedded, and it is denoted by $V(S,\pi)$, where $S\in\pi$. The pair $(N,V)$ gives rise to a partition function form game (Thrall and Lucas 1963). This is our primitive concept of a game in the paper. 

The coarsest partition, i.e., the partition $\{N\}$ is the efficient partition if $V(N,\{N\})>\sum\limits_{S\in\pi} V(S,\pi)$, for all $\pi\in \Pi.$ In this paper we will deal with games where this property holds. A vector of individual payoffs $z=(z_1,z_2,\ldots, z_n)$ is a feasible allocation of $V(N,\{N\})$ if $\sum\limits_{i\in N}z_i=V(N,\{N\})$. 

Take a coalition $S$ and denote by $\Pi_{S}$ the set of all partitions of $N$ that contain $S$. Denote by $h_{n,S}$ a probability distribution over $\Pi_S$. So for $\pi_S\in\Pi_S$, the value of $h_{n,S}(\pi_S)$ denotes the probability assigned by $S$ to partition $\pi_S$. We can then define a game $(N,V^h)$ where $V^h(S)=\sum\limits_{\pi_S\in\Pi_S}h_{n,S}(\pi_S)V(S,\pi_S).$ The worth of $N$ in this game is identical to that of $(N,V)$. Coalition $S$ objects to (or blocks) feasible allocation $z$ of $V(N,\{N\})$ in $(N,V^h)$ if $V^h(S)>\sum\limits_{i\in S}z_i.$ The core of $(N,V^h)$ is the set of all feasible allocations of $V(N,\{N\})$ that no coalition $S$ can block given  $h_{n,S}$.

We are interested in finding environments under which the above defined core is non-empty. To this end we will work with games where coalition formation creates monotonic -positive or negative- externalities.  These games are very often encountered in the literature. As in Hafalir (2007), \textit{positive externalities} arise if for any partition $\pi'$ of $N/\{S\cup S' \cup S''\}$, where $S,S',S''$ mutually disjoint coalitions, we have $V(S,\pi'\cup \{S'\cup S'', S\})>V(S,\pi'\cup \{S', S'', S\})$. Likewise, \textit{negative externalities} arise if $V(S,\pi'\cup\{S'\cup S'', S\})<V(S,\pi'\cup \{S', S'', S\})$. Hence, under positive externalities the merging of outside coalitions benefits $S$; the opposite holds under negative externalities.
%With a slight abuse of notation,\footnote{The reason why we index by $n$ will become evident later on.} denote by $\pi_{\gamma_n}$ the partition that includes $S$ and all outsiders as singletons; and by $\pi_{\delta_n}$ the partition that includes $S$ 

%Consider first $S_2$. We have
%$$\frac{v^3(N)}{3}=\frac{\langle x^N,u(x^N)\rangle}{3}\geq \frac{\langle 2x_{S_2},u(2x_{S_2})\rangle}{3}\geq \frac{\langle x_{S_2},u(2x_{S_2})\rangle}{2}= \frac{v^{f_3}(S_2)}{2}$$
To derive our results we impose symmetry (in the primitive game).

\vsp\noi \textit{Remark} \textit{$(N,V)$ is symmetric, i.e., for every partition $\pi$ of $N$ and $S,T\in \pi$, $V(S,\pi)=V(T,\pi)$ whenever $|S|=|T|$, where $|\cdot|$ denotes the number of elements of a set.}

\vsp The above remark will be adopted throughout the paper and we will not refer to it again.

The analysis will utilize induction in the number of players in the game. So it will be useful to write the worth of coalition $S$ as $V^{h_n}(S)$. In particular we will write $V^{\gamma_n}(S)$ to denote the worth of $S$ under the distribution that gives probability one to the scenario that all players outside $S$ form singleton coalitions ($\gamma$-conjecture); and $V^{\delta_n}(S)$ to denote the worth of $S$ under the distribution that gives probability one to the scenario that all players outside $S$ form one coalition ($\delta$-conjecture).
\section{Results}
\subsection{Positive externalities}
We will establish a certain pattern on the probabilistic beliefs of $S$ across games with different number of players. Namely, we will "tie" the beliefs of $S$ in a game with $n$ players to its beliefs in a game with $n+1$ players. This connection will help us in our induction argument: given the pattern, if $S$ does not deviate in a game with a certain number of players, it will also not deviate in a game with a larger number of players. 

To begin, notice first that positive externalities imply that $V^{h_n}(S)\in[V^{\gamma_n}(S),V^{\delta_n}(S)]$.
Given this, take an arbitrary distribution $h_{n,S}$ and denote by $\tilde{h}_{n+1,S}$ a distribution satisfying

\be\label{Mar2}V^{\tilde{h}_{n+1}}(S)=\frac{n}{n+1}V^{h_n}(S)\ee

\noi Whether such an $\tilde{h}_{n+1,S}$ exists is elaborated below (Propositions 1 and 2). Define  the set of distributions
\be\label{admset}B_{h_{n,S}}=\{h_{n+1,S}: V^{h_{n+1}}(S)\leq V^{\tilde{h}_{n+1}}(S)\}\ee

\vspace{0.1cm}\noi The above set is non-empty. For example, it contains the following non-empty set

$$B'_{h_{n,S}}=\{h_{n+1,S}: h_{n+1,S}(\pi_S)\leq \tilde{h}_{n+1,S}(\pi_S), \hspace{0.1cm} \mbox{for}\hspace{0.1cm} |\pi_S|\in\{2,3,\ldots,n-s\}\},$$

\vspace{0.1cm} where $s=|S|.$ Compared to $\tilde{h}_{n+1,S}$, any $h_{h+1,S}\in B'_{h_{n,S}}$ assigns uniformly lower probabilities to the most favorable partitions for $S$ and -thus- higher probability to the most unfavorable partition (recall we are in the case of positive externalities). Hence,  for any such $h_{n+1,S}$ the inequality $V^{h_{n+1}}(S)\leq V^{\tilde{h}_{n+1}}(S)$ must hold.

We will focus on the following beliefs of $S$, which we will define recursively w.r.t. the number of players in the game.

\vsp \noi \textbf{Definition 1} \textit{Consider the following beliefs:}
\begin{enumerate}[(i)]
	\item \textit{Let $n=3.$  Then $h_{3,S}$ is arbitrary.} 
	\item Let $n=4$. Then $h_{4,S}$ is restricted to be an element of $B_{h_{3,S}}$. 
	\item Let $n>4$ and assume that $h_{m, S}$ has been defined for all $m\leq n-1$. Then $h_{n,S}$ is restricted to be  an element of $B_{h_{n-1,S}}$.
\end{enumerate}
\noi \textit{The above beliefs will be called admissible.}
\vsp The use of the above will be clarified in Proposition 1.

\vsp We next define a closely related set of beliefs. Consider a distribution $h_{n,S}$ such that $V^{h_n}(S)\geq V^{\gamma_{n+1}}(S)$. Denote by ${\cal{R}}_{n,S}$ the set of all such distributions. Denote by $\tilde{h}_{n,S}$ a probability distribution satisfying 

\be\label{Mar22}V^{\tilde{h}_{n+1}}(S)=\frac{n}{n+1}V^{h_n}(S),\hspace{0.15cm}\mbox{where}\hspace{0.15cm} h_{n,S}\in {\cal{R}}_{n,S}\ee

Define finally the set

$${\cal{B}}^{\cal{R}}_{h_{n,S}}=\{h_{n+1,S}: V^{h_{n+1}}(S)\leq V^{\tilde{h}_{n+1}}(S)\},\hspace{0.13cm}\mbox{where}\hspace{0.2cm}h_{n,S}\in {\cal{R}}_{n,S}$$

\vsp The use of the above will be clarified in Proposition 2.

\vsp \noi \textbf{Definition 2} \textit{Consider the following beliefs:}
\begin{enumerate}[(i)]
	\item \textit{Let $n=3.$  Then $h_{3,S}$ is arbitrary.} 
	\item Let $n=4$. Then $h_{4,S}$ is restricted to be an element of $B^{\cal{R}}_{h_{3,S}}$. 
	\item Let $n>4$ and assume that $h_{m, S}$ has been defined for all $m\leq n-1$. Then $h_{n,S}$ is restricted to be  an element of $B^{\cal{R}}_{h_{n-1,S}}$.
\end{enumerate}
\noi \textit{The above beliefs will be called ${\cal{R}}$-admissible.}

\vsp We will break the analysis into two cases. In the first case we will assume that $V^{\gamma_{n+1}}(S)\leq \frac{n}{n+1}V^{\gamma_n}(S)$; in the second we will assume that the complementary case holds, i.e.,
$V^{\gamma_{n+1}}(S)>\frac{n}{n+1}V^{\gamma_n}(S)$. The first case will utilize admissible beliefs and the second will utilize ${\cal{R}}$-admissible beliefs.

Our first result is as follows.

\vsp\noi \textbf{Proposition 1} \textit{Let $V^{\gamma_{n+1}}(S)\leq \frac{n}{n+1}V^{\gamma_n}(S)$. If $S$ has admissible beliefs for any $|S|$, and if $h_{3,S}(1)$ is sufficiently low for $|S|=1$, then the core of $(N,V^h)$ is non-empty.}

\vsp\noi \textbf{Proof} The proof will use induction on $n\geq 3$.

\vspace{0.2cm}\noi \textit{Base of induction}: Consider a game with 3 players. We will show that the equal split allocation, i.e, the allocation that gives player $i$ the amount $z_i=\frac{v(N,\{N\})}{3}$, is a core allocation. I.e. we will show that

\be\label{maritsa1111}|S|\frac{V(N,\{N\})}{3}\geq V^{h_{n}}(S), \hspace{0.2cm}\mbox{for}\hspace{0.15cm}|S|\in\{1,2\}\ee

\vspace{0.15cm} Consider first a coalition $S$ with $|S|=2$. Take the partition $\pi_S=\{S,T\}$ of $N$, where $T=N\setminus S$, and so $|T|<|S|.$ Then using the equal split allocation, we have
$$\sum\limits_{i\in S}z_i=2\frac{v(N,\{N\})}{3}> 2\frac{V(S,\pi_S)+V(T,\pi_S)}{3}>2\frac{V(S,\pi_S)+V(S,\pi_S)/2}{3}= V(S,\pi_S),$$

where the first inequality uses the efficiency of partition $\{N\}$ and the second inequality is due to the property that under positive externalities the per member payoff of a small coalition is higher than the per member payoff of a larger coalition (Yi 1997, condition P.2, page 220).

Take now a coalition $S$ with $|S|=1.$ Recall that positive externalities imply $V^{h_3}(S)\in[V^{\gamma_3}(S), V^{\delta_3}(S)]$. Notice that $V^{\gamma_3}(S)$, which corresponds to $S$'s belief that the two outsiders stay separate with probability one, or form one coalition with probability zero, cannot exceed $\frac{v(N,\{N\})}{3}$, i.e., the per player efficient payoff.  Hence under this belief, $S$ does not deviate. 

As the probability that the outsiders form one coalition increases from zero, the payoff of $S$ increases monotonically towards $v^{\delta_3}(S)$. There are two possible cases: 

\begin{enumerate}[(i)]
\item  $v^{\delta_3}(S)\leq \frac{v(N,\{N\})}{3}$: then $S$ does not deviate irrespective of its beliefs.

\item $v^{\delta_3}(S)> \frac{v(N,\{N\})}{3}$: then $S$ does not deviate if the probability attached to the scenario that the outsiders form one coalition, i.e., $h_{3,S}(1)$, does not exceed a critical value.
\end{enumerate}

\noi The current model  cannot tell us whether (i) or (ii)  holds. So, we can only deduce that $S$ does not deviate if the probability that it assigns to the scenario of a sole outside coalition is low enough. All the above complete the Base of induction.

\vsp\noi {\it{Induction hypothesis}}: Assume that in a game with $n$ players the equal split allocation, i.e, the allocation that gives player $i$ the amount $z_i=\frac{v(N,\{N\})}{n}$, is a core allocation. I.e.,

\be\label{maritsa111}|S|\frac{V(N,\{N\})}{n}\geq V^{h_{n}}(S), \hspace{0.2cm}\mbox{for}\hspace{0.15cm}|S|\in\{1,2,\ldots,n-1\}\ee

\vsp\noi {\it{Induction step}}: We will show that in a game with $n+1$ players, the equal split allocation, i.e, the allocation that gives player $i$ the amount $z_i=\frac{v(N,\{N\})}{n+1}$, is a core allocation. I.e., we will show that

\be\label{march9a11}|S|\frac{V(N,\{N\})}{n+1}\geq V^{h_{n+1}}(S), \hspace{0.2cm}\mbox{for}\hspace{0.15cm} S\in\{1,2,\ldots,n\}\ee

\vsp By the Induction hypothesis and the observation that the worth of the grand coalition is a weakly increasing function\footnote{This holds if we assume that the addition of an extra player either creates value or, in the opposite case, that extra player can become inactive -without affecting the worth of the grand coalition.} of $|N|$, it suffices to show that  

\be\label{maritsa3}(n+1)V^{h_{n+1}}(S)\leq nV^{h_n}(S)\ee

\vsp\noi Since\footnote{As noted, the opposite case,  $V^{\gamma_{n+1}}(S)> \frac{n}{n+1}V^{\gamma_n}(S)$, will be examined in Proposition 2.} $V^{\gamma_{n+1}}(S)\leq \frac{n}{n+1}V^{\gamma_n}(S)$, we have the following cases:

\vspace{0.3cm} \noi\textit{(i) $V^{\delta_{n+1}}(S)<\frac{n}{n+1}V^{\gamma_n}(S)$}

\vspace{0.25cm} Then we have the following picture of Figure 1:

\vspace{-0.6cm}
\begin{center}
	\begin{figure}[htpb]
		\caption{\textit{}}
		\vsp
		\label{fig:hotel9}
		\begin{center}
			\begin{tikzpicture}[xscale=11]
			\draw[-] [ thick] (0,0)--(1,0);
			\draw [thick](0,-0.145) node[below]{\scriptsize{$V^{\gamma_{n+1}}(S)$}}-- (0,0.1);
			\draw [thick](1,-0.1) node[below]{\scriptsize{$\frac{n}{n+1}V^{\delta_{n}}(S)$}}-- (1,0.1);
			\draw [thick](0.36,-0.145) node[below]{\scriptsize{$V^{\delta_{n+1}}(S)$}}-- (0.36,0.1);
			
			\draw [thick](0.71,-0.13) node[below]{\scriptsize{$\frac{n}{n+1}V^{\gamma_n}(S)$}}-- (0.71,0.1);

			\node [black] at (0.855,0.5) {\scriptsize{$\underbrace{\hspace{0.84cm}\frac{n}{n+1}V^{h_n}(S)}$}\hspace{0.9cm}};
			
			%\node [black] at (0.4,-1.6) {$x$};
			
			%\draw (0.2,-0.3)--(0.4,-0.3){\underbrace{x}};
			
			%\draw [decorate] ([yshift=-5mm]) --node[below=3mm]{$n_1$} ([yshift=-5mm]g3.east);

			%\draw [thick](0.76,+0.1) node[above]{Επιχείρηση 2}-- (0.76,0.1);
			%\draw [thick](0.25,+0.1) node[above]{Επιχείρηση 1}-- (0.25,0.1);
			%\draw [thick](0.5,-0.425) node[below]{Καταναλωτής $x$}-- (0.5,-0.425);
			
			\end{tikzpicture}
		\end{center}
	\end{figure}
\end{center}

Hence (\ref{maritsa3}) holds automatically.

\vspace{0.4cm}\noi \textit{(ii) $\frac{n}{n+1}V^{\gamma_n}(S)<V^{\delta_{n+1}}(S)<\frac{n}{n+1}V^{\delta_{n}}(S)$}

\vspace{0.25cm} Then we have the following picture of Figure 2:

\vspace{-0.1cm}
\begin{center}
	\begin{figure}[htpb]
		\caption{\textit{}}
		\vsp
		\label{fig:hotel1}
		\begin{center}
			\begin{tikzpicture}[xscale=11]
			\draw[-] [ thick] (0,0)--(1,0);
			\draw [thick](0,-0.145) node[below]{\scriptsize{$V^{\gamma_{n+1}}(S)$}}-- (0,0.1);
			\draw [thick](1,-0.1) node[below]{\scriptsize{$\frac{n}{n+1}V^{\delta_{n}}(S)$}}-- (1,0.1);
			\draw [thick](0.36,-0.145) node[below]{\scriptsize{$\frac{n}{n+1}V^{\gamma_{n}}(S)$}}-- (0.36,0.1);
			
			\draw [thick](0.71,-0.13) node[below]{\scriptsize{$V^{\delta_{n+1}}(S)$}}-- (0.71,0.1);

			\node [black] at (0.66,0.5) {\scriptsize{$\underbrace{\hspace{5cm}\frac{n}{n+1}V^{h_n}(S)}\hspace{-0.6cm}$}\hspace{6.9cm}};
			
		%	\node [black] at (0.4,-1.6) {$x$};
			
			%\draw (0.2,-0.3)--(0.4,-0.3){\underbrace{x}};
			
			%\draw [decorate] ([yshift=-5mm]) --node[below=3mm]{$n_1$} ([yshift=-5mm]g3.east);

			%\draw [thick](0.76,+0.1) node[above]{Επιχείρηση 2}-- (0.76,0.1);
			%\draw [thick](0.25,+0.1) node[above]{Επιχείρηση 1}-- (0.25,0.1);
			%\draw [thick](0.5,-0.425) node[below]{Καταναλωτής $x$}-- (0.5,-0.425);
			
			\end{tikzpicture}
		\end{center}
	\end{figure}
\end{center}

\noi By the above we can see that for each $h_{n,S}$ either (a) or (b) holds: 

\noi (a) there exists a non-unique $\tilde{h}_{n+1,S}$ such that $V^{\tilde{h}_{n+1}}(S)=\frac{n}{n+1}V^{h_n}(S)$ and $V^{h_{n+1}}(S)\leq V^{\tilde{h}_{n+1}}(S)$ for all $h_{n+1,S}\in B_{h_{n,S}}$. For all these distributions (\ref{maritsa3}) holds.

\vspace{0.2cm}\noi (b) Expression (\ref{maritsa3}) holds automatically.

\vspace{0.4cm} \noi\textit{(iii) $V^{\delta_{n+1}}(S)>\frac{n}{n+1}V^{\delta_{n}}(S)$}

\vspace{0.2cm} Then we have Figure 3:

\vspace{-0.6
	cm}
\begin{center}
	\begin{figure}[htpb]
		\caption{\textit{}}
		\vsp
		\label{fig:hotel2}
		\begin{center}
			\begin{tikzpicture}[xscale=11]
			\draw[-] [ thick] (0,0)--(1,0);
			\draw [thick](0,-0.145) node[below]{\scriptsize{$V^{\gamma_{n+1}}(S)$}}-- (0,0.1);
			\draw [thick](1,-0.1) node[below]{\scriptsize{$V^{\delta_{n+1}}(S)$}}-- (1,0.1);
			\draw [thick](0.36,-0.145) node[below]{\scriptsize{$\frac{n}{n+1}V^{\gamma_n}(S)$}}-- (0.36,0.1);
			
			\draw [thick](0.71,-0.13) node[below]{\scriptsize{$\frac{n}{n+1}V^{\delta_{n}}(S)$}}-- (0.71,0.1);

			\node [black] at (0.53,0.5) {\scriptsize{$\underbrace{\hspace{0.84cm}\frac{n}{n+1}V^{h_n}(S)}$}\hspace{0.9cm}};
			
			%\node [black] at (0.4,-1.6) {$x$};
			
			%\draw (0.2,-0.3)--(0.4,-0.3){\underbrace{x}};
			
			%\draw [decorate] ([yshift=-5mm]) --node[below=3mm]{$n_1$} ([yshift=-5mm]g3.east);

			%\draw [thick](0.76,+0.1) node[above]{Επιχείρηση 2}-- (0.76,0.1);
			%\draw [thick](0.25,+0.1) node[above]{Επιχείρηση 1}-- (0.25,0.1);
			%\draw [thick](0.5,-0.425) node[below]{Καταναλωτής $x$}-- (0.5,-0.425);
			
			\end{tikzpicture}
		\end{center}
	\end{figure}
\end{center}

\noi By the above we can see the following: For each distribution $h_{n,S}$ there exists a non-unique distribution $\tilde{h}_{n+1,S}$ such that $V^{\tilde{h}_{n+1}}(S)=\frac{n}{n+1}V^{h_n}(S)$ and
$V^{h_{n+1}}(S)\leq V^{\tilde{h}_{n+1}}(S)$, for $h_{n+1,S_1}\in B_{h_{n,S}}$. 

Hence under admissible beliefs, (\ref{maritsa3}) holds. This concludes the Induction step. \qed

\vsp\noi Admissible beliefs essentially tells us that, as the number of outside players increases, coalition $S$ assigns higher probability weights to partitions consisting of few members. Combining this with positive externalities, allows our induction argument to work. The justification of admissible beliefs can be based on coordination issues: as the number of outside players increases, the  members of $S$ believe that it is more difficult for the outsiders to coordinate into forming coalitions containing many players.

\vsp We now turn to the case where $V^{\gamma_{n+1}}(S)>\frac{n}{n+1}V^{\gamma_n}(S)$. For this case we will utilize ${\cal{R}}$-admissible beliefs.

\vsp\noi \textbf{Proposition 2} \textit{Let $V^{\gamma_{n+1}}(S)>\frac{n}{n+1}V^{\gamma_n}(S)$. If $S$ has ${\cal{R}}$-admissible beliefs for any $|S|$, and if $h_{3,S}(1)$ is sufficiently low for $|S|=1$, then the core of $(N,V^h)$ is non-empty.}

\vsp\noi \textbf{Proof} Similarly to the proof of Proposition 1 until expression (\ref{maritsa3}).
Then we have the following cases to examine.

(i) $V^{\delta_{n+1}}(S)\leq \frac{n}{n+1}V^{\delta_{n}}(S)$

\vspace{0.2cm} Then Figure 4 is in order:

\vspace{-0.6
	cm}
\begin{center}
	\begin{figure}[htpb]
		\caption{\textit{}}
		\vsp
		\label{fig:hotel4}
		\begin{center}
			\begin{tikzpicture}[xscale=11]
			\draw[-] [ thick] (0,0)--(1,0);
			\draw [thick](0,-0.145) node[below]{\scriptsize{$\frac{n}{n+1}V^{\gamma_{n}}(S)$}}-- (0,0.1);
			\draw [thick](1,-0.1) node[below]{\scriptsize{$\frac{n}{n+1}V^{\delta_{n}}(S)$}}-- (1,0.1);
			\draw [thick](0.36,-0.145) node[below]{\scriptsize{$V^{\gamma_{n+1}}(S)$}}-- (0.36,0.1);
			
			\draw [thick](0.71,-0.13) node[below]{\scriptsize{$V^{\delta_{n+1}}(S)$}}-- (0.71,0.1);

			\node [black] at (0.53,0.5) {\scriptsize{$\underbrace{\hspace{0.84cm}V^{h_{n+1}}(S)}$}\hspace{0.9cm}};
			
			%\node [black] at (0.4,-1.6) {$x$};
			
			%\draw (0.2,-0.3)--(0.4,-0.3){\underbrace{x}};
			
			%\draw [decorate] ([yshift=-5mm]) --node[below=3mm]{$n_1$} ([yshift=-5mm]g3.east);

			%\draw [thick](0.76,+0.1) node[above]{Επιχείρηση 2}-- (0.76,0.1);
			%\draw [thick](0.25,+0.1) node[above]{Επιχείρηση 1}-- (0.25,0.1);
			%\draw [thick](0.5,-0.425) node[below]{Καταναλωτής $x$}-- (0.5,-0.425);
			
			\end{tikzpicture}
		\end{center}
	\end{figure}
\end{center}

Then for each $h_{n,S}\in{\cal{R}}_{n,S}$, either (a) or (b) holds:

\vspace{0.1cm} (a) there exists a non-unique  distribution $\tilde{h}_{n+1,S}$ such that $V^{\tilde{h}_{n+1}}(S)=\frac{n}{n+1}V^{h_n}(S)$ and $V^{h_{n+1}}(S)\leq V^{\tilde{h}_{n+1}}(S)$ for all $h_{n+1,S}\in{\cal{B}}^{R}_{h_{n,s}}$.

\vspace{0.1cm} (b) Expression (\ref{maritsa3}) holds automatically.

\vsp (ii) $\frac{n}{n+1}V^{\delta_n}(S)<V^{\delta_{n+1}}(S)$

\vspace{0.2cm} Then Figure 5 is in order:

\vspace{-0.6
	cm}
\begin{center}
	\begin{figure}[htpb]
		\caption{\textit{}}
		\vsp
		\label{fig:hotel6}
		\begin{center}
			\begin{tikzpicture}[xscale=11]
			\draw[-] [ thick] (0,0)--(1,0);
			\draw [thick](0,-0.145) node[below]{\scriptsize{$\frac{n}{n+1}V^{\gamma_n}(S)$}}-- (0,0.1);
			\draw [thick](0.71,-0.13) node[below]{\scriptsize{$\frac{n}{n+1}V^{\delta_n}(S)$}}-- (0.71,0.1);
			
			\draw [thick](0.36,-0.145) node[below]{\scriptsize{$V^{\gamma_{n+1}}(S)$}}-- (0.36,0.1);
			
			\draw [thick](1,-0.1) node[below]{\scriptsize{$V^{\delta_{n+1}}(S)$}}-- (1,0.1);

			\node [black] at (0.6,0.5) {\scriptsize{$\underbrace{\hspace{6.4cm}V^{h_{n+1}}(S)}$}\hspace{0.1cm}};
			
			%\node [black] at (0.4,-1.6) {$x$};
			
			%\draw (0.2,-0.3)--(0.4,-0.3){\underbrace{x}};
			
			%\draw [decorate] ([yshift=-5mm]) --node[below=3mm]{$n_1$} ([yshift=-5mm]g3.east);

			%\draw [thick](0.76,+0.1) node[above]{Επιχείρηση 2}-- (0.76,0.1);
			%\draw [thick](0.25,+0.1) node[above]{Επιχείρηση 1}-- (0.25,0.1);
			%\draw [thick](0.5,-0.425) node[below]{Καταναλωτής $x$}-- (0.5,-0.425);
			
			\end{tikzpicture}
		\end{center}
	\end{figure}
\end{center}

Then for each $h_{n,S}\in{\cal{R}}_{n,S}$, there exists a non-unique  $\tilde{h}_{n+1,S}$ such that $V^{\tilde{h}_{n+1}}(S)=\frac{n}{n+1}V^{h_n}(S)$ and $V^{h_{n+1}}(S)\leq V^{\tilde{h}_{n+1}}(S)$ for all $h_{n+1,S}\in{\cal{B}}^{R}_{h_{n,s}}$.

Hence, under ${\cal{R}}$-admissible beliefs (\ref{maritsa3}) holds. This concludes the Induction step. \qed

\vsp  ${\cal{R}}$-admissible beliefs are justified in a way similar to admissible beliefs. 

We note that the need to restrict attention to ${\cal{R}}$-admissible beliefs in Proposition 2 is merely technical: ${\cal{R}}$-admissible beliefs allow for the induction step to be executed in the specific case where the inequality $V^{\gamma_{n+1}}(S)>\frac{n}{n+1}V^{\gamma_n}(S)$ holds.

\subsection{Negative externalities}
In this part we discuss the case of negative externalities. The analysis is the mirror of the analysis under positive externalities, so we will simply point out the differences between the two cases.

Notice first that under negative  externalities we have that $V^{h_n}(S)\in[V^{\delta_n}(S),V^{\gamma_n}(S)]$, as the merging of outside coalitions hurts $S$ (i.e., the inverse compared to positive externalities). 

Following the steps of the previous subsection, we may again define a distribution $\tilde{h}_{n+1,S}$ that satisfies (\ref{Mar2}) and also define a set like (\ref{admset}). However, due to negative externalities, a non-empty subset of $B_{h_{n,S}}$ will now be of the form

$$B''_{h_{n,S}}=\{h_{n+1,S}: h_{n+1,S}(\pi_S)\geq \tilde{h}_{n+1,S}(\pi_S), \hspace{0.1cm} \mbox{for}\hspace{0.1cm} |\pi_S|\in\{1,2,\ldots,n-s-1\}\}$$

Compared to $\tilde{h}_{n+1,S}$, any $h_{h+1,S}\in B''_{h_{n,S}}$ assigns uniformly higher probabilities to the most unfavorable partitions and thus lower probability to the most favorable partition. Hence,  for any such $h_{n+1,S}$ the inequality $V^{h_{n+1}}(S)\leq V^{\tilde{h}_{n+1}}(S)$ holds.

To show the non-emptiness of core, we may again use induction and also tie the beliefs of $S$ in a game with $n$ players to its beliefs in a game with $n+1$ players. However, instead of having admissible beliefs, which assign relatively lower probabilities to partitions with few members as $n$ increases, we now require the opposite: as $n$ increases, coalition $S$ assigns relatively low probabilities to partitions with relatively many members. 

Regarding the induction per se, the analysis is similar to that of Propositions 1 and 2, with the difference that when we examine the Induction step and in particular expression  (\ref{maritsa3}), coalitional beliefs will be guided by the principles stated in the previous paragraphs.

\section*{References}

\begin{enumerate}[1.]
	
%	\item Acemoglu D., M.K. Jensen (2013) Aggregate comparative statics, Games and Economic Behavior,  81, 27-49.

	\item Aumann R. (1959) Acceptable points in general cooperative n-person games, in: Tucker, Luce (eds.), Contributions to the theory of games IV, Annals of Mathematics Studies 40, Princeton University Press, Princeton.		
	
	%\item Bloch F. (1996) Sequential formation of coalitions in games with externalities and fixed payoff division, Games and Economic  Behavior, 14, 90-123.
	
	\item Bloch F., A. Nouweland (2014) Expectation formation rules and the core of partition function games, Games and Economic  Behavior 88, 359-353.
	
	\item Chander P., H. Tulkens (1997) A core of an economy with multilateral environmental externalities, International Journal of Game Theory 26, 379-401.
	
	%\item Chander P. (2007) The $\gamma$-core and coalition formation, International Journal of Game Theory, 35, 539-556.
	
	%\item Chander P., M. Wooders (2012) Subgame-perfect cooperation in an extensive game, Working Paper.
	
	%\item Corch$\acute{\mbox{o}}$n L.C. (1994) Comparative statics for aggregative games. The strong concavity case, Mathematical Social Sciences, 28, 151-165.
	
	%\item Cornes, R., Hartley, R. (2012)  Fully aggregative games, Economics Letters, 116, 631-633
	
	\item Currarini S., M. Marini (2003) A sequential approach to the characteristic function and the core in games with externalities, Kara A., Sertel, M. (eds.), Advances in Economic Design. Springer Verlag, Berlin.
	
	%\item Gabszewicz J., M. Marini, O. Tarola (2016) Core existence in vertically differentiated markets, Economics Letters, 149, 28-32.
	
	\item Hafalir I. E. (2007) Efficiency in coalition games with externalities, Games and Economic Behavior 61, 242-258.
	
	\item Hart S., M. Kurz (1983) Endogenous formation of coalitions, Econometrica 51, 1047-1064.

	%\item Helm C. (2001) On the existence of a cooperative solution for a cooperative game with externalities, International Journal of Game Theory, 30, 141-146.	

	\item Huang C.Y., T. Sjostrom (2003) Consistent solutions for cooperative games with externalities, Games and Economic Behavior 43, 196-213.
	
	\item K\'{o}czy L. (2007) A recursive core for partition function form games, Theory and Decision 63, 41-51.
	
	\item K\'{o}czy L. (2018) Partition function form games: Coalitional games with externalities, Springer. 
	
	%\item Lardon A. (2010) Convexity of Bertrand oligopoly TU-games with differentiated products. Working Paper.			
	
%	\item Lardon A. (2012) The $\gamma$-core in Cournot oligopoly TU-games with capacity constraints, Theory and Decision, 72, 387-411.
	
	\item Lekeas P., G. Stamatopoulos (2014) Cooperative oligopoly games with boundedly rational firms, Annals of Operations Research 223, 255-272.
	
	\item Marini M. (2013) The sequential core of an economy with environmental externalities, Environmental Sciences 1, 79-82.			
	
%	\item Martimort D., L. Stole (2010) Aggregate representations of aggregate games, Working Paper.

	%\item McKelvey R., T. R. Palfrey (1995) Quantal response equilibria for normal form games, Games and Economic Behavior 10, 6-38.
	
	\item Nax H.H. (2014) A note on the core of TU-cooperative games with multiple membership externalities, Games 5, 191-203.
	
	\item Stamatopoulos G. (2021) On the core of economies with multilateral environmental externalities, Journal of Public Economic Theory 23, 158-171.
	
%	\item Norde, H., K. H. P. Do, S. Tijs (2002)  Oligopoly games with and without transferable technologies, Mathematical Social Sciences, 43, 187-207.
	
%	\item Rajan R. (1989) Endogenous coalition formation in cooperative oligopolies, International Economic Review, 30, 863-876.
	
	%\item Yi S.S. (1997). Stable coalition structures with externalities, Games and Economic Behavior, 20, 201-237.
	
	\item Thrall R.M., W.F. Lucas (1963) $n$-Person games in partition function form, Naval Research Logistics Quarterly 10, 281-298.
	
	\item Yi S.S. (1997) Stable coalition structures with externalities, Games and Economic Behavior 20, 201-237.
	
%	\item Zhao  J. (1999) A $\beta$-core existence result and its application to oligopoly markets, Games and Economic Behavior, 27, 153-168.
	
\end{enumerate}
\end{document}